\documentclass[prl,showpacs,amsfonts,amssymb,floats,twocolumn,superscriptaddress,aps]{revtex4}

\usepackage[final,dvips]{epsfig}
\newcommand{\pd}{{\phantom{\dagger}}}

\begin{document}

\title[]
{Electron Transfer in Donor-Acceptor Systems: Many-Particle Effects and
Influence of Electronic Correlations
}

\author{Sabine Tornow}

\affiliation{\mbox{Theoretische Physik III, Elektronische Korrelationen und
Magnetismus, Universit\"at Augsburg,
86135 Augsburg, Germany}}
\author{Ning-Hua Tong}
\affiliation{\mbox{Institut f\"ur Theorie der Kondensierten Materie,
Universit\"at Karlsruhe,
76128 Karlsruhe, Germany}}
\author{Ralf Bulla}
\affiliation{\mbox{Theoretische Physik III, Elektronische Korrelationen und
Magnetismus, Universit\"at Augsburg,
86135 Augsburg, Germany}}
\date{\today}

\begin{abstract}
We investigate electron transfer processes in donor-acceptor
systems with a coupling of the electronic degrees of freedom to a
common bosonic bath. The model allows to study many-particle
effects and the influence of the local Coulomb interaction $U$
between electrons on donor and acceptor sites. Using the
non-perturbative numerical renormalization group approach we find
distinct differences between the electron transfer characteristics
in the single- and two-particle subspaces. We calculate the
critical electron-boson coupling $\alpha_{\rm c}$ as a function of
$U$ and show results for density-density correlation functions in
the whole parameter space. The possibility of many-particle
(bipolaronic) and Coulomb-assisted transfer is discussed.
\end{abstract}


\maketitle

{\em Introduction} -
Electron transfer (ET) is a fundamental process in chemistry,
biology and physics, for example in corrosion of metals, charge transfer
in semiconductors, enzymatic activities, cell metabolism, and
photosynthesis \cite{Jortner,May}.
The characterization of ET processes in bio-molecules is an important
step towards an understanding of the biological function of
many proteins and towards the possible construction of bio-molecular
electronic devices or biosensors.
Theoretical investigations of ET processes typically start from a two-site
model for the electronic degrees of freedom at the donor and
acceptor sites which are coupled via a tunneling matrix
element $t$. The correlated dynamics of electrons and vibronic
modes (phonons) is essential for the ET characteristics. If the
phonons are treated classically, one arrives at the
Marcus theory \cite{Marcus}. A quantum mechanical treatment of the
phononic degrees of freedom \cite{Garg} results in models related to the
spin-boson model, in which the phonons
are modeled by an infinite set of harmonic
oscillators with a continuous spectral density 
$J(\omega)$ \cite{Leggett,Weiss}.

The following two limiting cases of ET processes are 
well understood. If the tunneling (or hopping)
matrix element $t$ is small we are in the limit of
nonadiabatic ET. In the opposite limit, where $t$ is large, the ET is
adiabatic and the Born-Oppenheimer theorem holds. (The time of the
electron moving from the donor to the acceptor is too short for
the vibronic modes to change their configuration.). 
Both limits are realized in ET processes in proteins
but of
particular interest are those parameters which lie in the crossover
regime. In this case, non-perturbative methods have to be applied
(see, for example, Ref.~\onlinecite{Egger}).

If the electronic part is treated in a one-particle picture it
reduces to two localized quantum states and can be modeled by a
two-level system or spin, leading to a description
in terms of the spin-boson model. In
many cases this may not be the adequate picture -- many-particle
effects and electron-electron interactions have to be taken into
account, for example if a more realistic modeling of the electronic
degrees of freedom is required
\cite{Ondrechen,Cox}, if more than one electron is transfered
simultaneously \cite{Zusman}, or for a proper description of
 exciton transfer \cite{Schulten1}.

\begin{figure}[b]
\vspace*{-0.4cm}
\epsfxsize=1.9in
\centerline{\epsffile{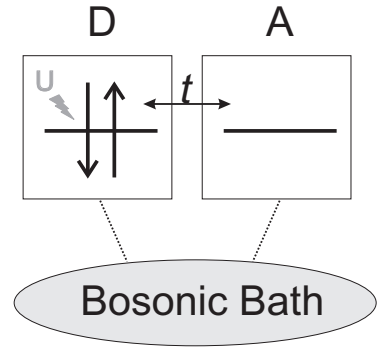}}
\caption{
Schematic view of the two-site electron-boson model
investigated in this paper.
The tunneling matrix element of electrons between donor (D) and
acceptor (A) sites is given by $t$; the parameter $U$ denotes
the local Coulomb interaction. Dissipation in
the electron transfer process is due to the coupling of the
electronic degrees of freedom to a common bosonic bath.
}
\label{fig:model}
\end{figure}

In this paper we propose a two-site electron-boson model to
investigate many-particle effects and the role of electron-electron
interactions. As sketched in Fig.~\ref{fig:model}, the model contains both
a local Coulomb interaction $U$ and the coupling to a common bosonic
bath.
Our approach goes beyond earlier work in which only the electronic
degrees of freedom involved in the ET transfer were considered
(see Refs.~\onlinecite{Schulten1,Ondrechen,Cox}) or the
coupling to the bosonic bath was treated within the spin-boson
model \cite{Egger} without considering 
many-particle effects and electronic correlations.

We discuss competition or cooperation of vibronic and electronic
effects (electron correlations) during the ET process by using the bosonic
numerical renormalization group (NRG) method \cite{BTV,BLTV} which is
non-perturbative and can be used in the whole parameter regime
between the adiabatic and nonadiabatic limits. We indeed find a 
significant difference between the ET characteristics in the
single-particle subspace (which can be mapped onto the spin-boson
model) and the two-particle subspace in which many-particle
effects have to be considered. The coupling to the bosonic bath
turns out to favor the formation of bipolarons, whereas a local
Coulomb repulsion tends to delocalize the electrons.

{\em Model and Method} -
The two-site electron-boson model is given by the following Hamiltonian:
\begin{eqnarray}
  H_{\rm eb}
&=& \sum_{\sigma,i={\rm A,D}} \varepsilon_i c^\dagger_{i\sigma} c^\pd_{i\sigma}
     - t \sum_\sigma \left( c^\dagger_{{\rm D}\sigma} c^\pd_{{\rm A}\sigma} +
                               c^\dagger_{{\rm A}\sigma} c^\pd_{{\rm D}\sigma}
                         \right)   \nonumber \\
    &+& U \sum_{i={\rm A,D}}  c^\dagger_{i\uparrow} c^\pd_{i\uparrow}
                              c^\dagger_{i\downarrow} c^\pd_{i\downarrow}
     + \sum_{n} \omega_{n} b_{n}^{\dagger} b^\pd_{n}
    \nonumber \\
    &+& (g_{\rm A} n_{\rm A} + g_{\rm D} n_{\rm D}) \sum_{n}
         \frac{\lambda_n}{2} \left(  b_{n}^{\dagger} + b^\pd_{n}  \right)
    \ .
\label{eq:ebm}
\end{eqnarray}
The operators $c^{(\dagger)}_{i\sigma}$ denote annihilation (creation)
operators for fermions with spin $\sigma$ on the donor ($i={\rm D}$)
and acceptor ($i={\rm A}$) sites; $n_{\rm A/D}$ is defined as
$n_i=\sum_\sigma c^{\dagger}_{i\sigma}c_{i\sigma}$.
The first three terms of the
Hamiltonian eq.~(\ref{eq:ebm}) correspond to the two-site Hubbard model
investigated in Ref.~\onlinecite{Ondrechen},
with $\varepsilon_i$ the on-site energies,
$t$ the hopping matrix element, and $U$ the local Coulomb interaction for
two electrons on either donor or acceptor sites.
This part of the model
can be easily extended to include non-local Coulomb correlations
between electrons on donor and acceptor sites;
 for simplicity, we restrict the discussion
here to local Coulomb terms only. The last two terms in
eq.~(\ref{eq:ebm}) describe the free bosonic bath and the coupling
between electrons and bosons, respectively. In the following we
set  $g_{\rm A}=1$ and $g_{\rm D}=-1$, assuming a symmetric
shift of the phonon displacements due to the electronic occupancy
at donor and acceptor sites \cite{May}.

In analogy to the spin-boson model \cite{Leggett,Weiss}, the coupling
of the electrons to the bath degrees of freedom is completely
specified by the bath spectral function
\begin{equation}
   J(\omega) = \pi \sum_{n}
\lambda_{n}^{2} \delta\left( \omega -\omega_{n} \right) \ .
\end{equation}
Here we assume an Ohmic bath, corresponding to the situation in many
proteins \cite{Schulten2}, with $J(\omega) = 2\pi\alpha\omega$,
$0<\omega<\omega_{\rm c}$, with $\alpha$ the dimensionless
coupling strength and $\omega_{\rm c}$ a cut-off which sets the
energy scale in the following ($\omega_{\rm c}=1$).

The Hamiltonian eq.~(\ref{eq:ebm}) conserves both particle number
$n_{\rm el}$ and total spin of the electrons.
The Hilbert space of the full model can therefore be divided
into subspaces labeled by $(Q,S_z)$, with
$Q=n_{\rm el}-2$ (so that $Q=0$ corresponds to half-filling) 
and $S_z$ the $z$-component
of the total spin.

Here we only consider the subspaces
$(Q,S_z) = (-1,1/2)$ and $(Q,S_z) = (0,0)$.
In the subspace $(-1,1/2)$, which we term the single-electron
subspace, the model eq.~(\ref{eq:ebm}) is equivalent to the
spin-boson model
\begin{equation}
H_{\rm sb}=-\frac{\Delta}{2}\sigma_{x}+\frac{\epsilon}{2}\sigma_{z}+
\sum_{n} \omega_{n}
     a_{n}^{\dagger} a_{n}
+\frac{\sigma_{z}}{2} \sum_{n}
    \lambda_{n}( a_{n} + a_{n}^{\dagger} ) \ ,
\label{eq:sbm}
\end{equation}
with
    $t = \Delta/2$,
    $\varepsilon_{\rm A} = \epsilon/2$, and
    $\varepsilon_{\rm D} = -\epsilon/2$. 
The two-electron subspace $(Q,S_z) =(0,0)$ is the main focus of the
calculations presented in this paper;
The electronic degrees of freedom in this subspace
can be represented by the four-dimensional
basis
\begin{equation}
   \vert i \rangle = \left\{
     \vert \uparrow\downarrow,0 \rangle,
     \vert \uparrow,\downarrow \rangle,
     \vert \downarrow,\uparrow \rangle,
     \vert 0,\uparrow\downarrow \rangle
                     \right\}   \ ,
\end{equation}
with the notation $\vert a,d \rangle$ describing the occupation
at the donor ($d$) and acceptor ($a$) sites. Consider
now the $4\times4$-matrix $M_{\rm eb}^{Q=0}=\langle i \vert H_{\rm eb}
\vert j \rangle$
($i,j=1,\ldots, 4$) with the matrix elements taken only with
respect to the electronic degrees of freedom. Introducing
the notation
\begin{equation}
    \hat{Y} = \sum_{n} \omega_{n} b_{n}^{\dagger} b^\pd_{n}
    \ \ , \ \
    \hat{X} = \sum_{n}
         \frac{\lambda_n}{2} \left(  b_{n}^{\dagger} + b^\pd_{n}  \right)
    \ ,
\end{equation}
we arrive at the matrix
\begin{equation}
   M_{\rm eb}^{Q=0}=
 \left( \begin{array}{cccc}
         \epsilon+ U +2 \hat{X} + \hat{Y} & -t & t & 0\\
         -t &  \hat{Y} & 0 & -t\\
         t & 0 &   \hat{Y} & t \\
         0 & -t & t &  -\epsilon+ U -2 \hat{X} + \hat{Y}
 \end{array}     \right) \ .
\end{equation}
The matrix $M_{\rm eb}^{Q=0}$
defines the starting point for our numerical calculations.

The technique we are using here, the bosonic NRG, has been
described in detail in Refs.~\onlinecite{BTV,BLTV} in the context of
the spin-boson model. The basic features of the bosonic NRG are as
follows: (i) the logarithmic discretization of the bath spectral
function $J(\omega)$ in intervals $[\Lambda^{-n+1},\Lambda^{-n}]$,
with $n=0,1,\ldots,\infty$ and $\Lambda>1$ the NRG discretization
parameter (all the results shown in this paper have been
calculated using $\Lambda=2$); 
within each of these intervals only one bosonic degree
of freedom is retained as a representative of the continuous set
of degrees of freedom. (ii) The mapping of the resulting
Hamiltonian onto a semi-infinite chain. (iii) The iterative
diagonalization of the chain-Hamiltonian via successively adding
one site to the chain.

The bosonic NRG has been shown to give very accurate results for
the spin-boson model \cite{BTV,BLTV}. One of its strengths is the
flexibility to handle a variety of models involving the coupling
of a small subsystem (here the electronic degrees of freedom at
donor and acceptor sites) to a bosonic bath. The application to
the two-electron subspace of the two-site electron-boson model is
therefore straightforward.

\begin{figure}[!t]
\epsfxsize=2.7in
\centerline{\epsffile{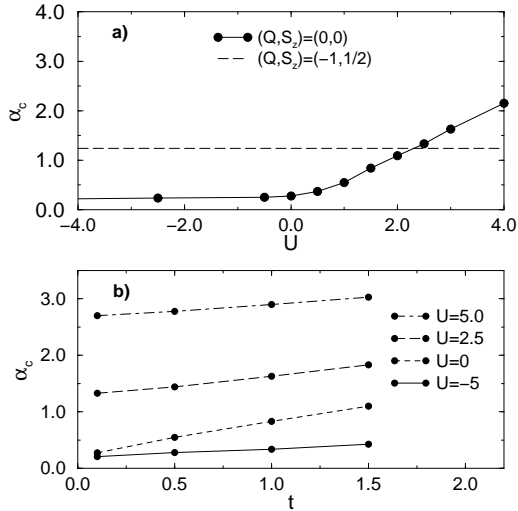}}
\vspace*{-0.4cm}
\caption{
Phase diagrams of the two-site electron-boson model in the
zero-bias case, $\epsilon=0$;
a) dependence of the critical
dissipation strength $\alpha_{\rm c}^{Q=0}$ and $\alpha_{\rm c}^{Q=-1}$ on the Coulomb interaction
$U$ for $t=0.1$ in the two-particle subspace (solid line)
and in the single-particle subspace (dashed line), respectively.
b) dependence of $\alpha_{\rm c}^{Q=0}$ on the hopping matrix element
$t$ for various values of $U$.
}
\label{fig:pd}
\vspace*{-0.4cm}
\end{figure}

{\em Results} - Let us first consider the zero-bias case
$\varepsilon_{\rm A}=\varepsilon_{\rm D}=0$ for temperature
$T=0$
\cite{finite}. In the single-electron subspace, this gives
$\epsilon =0$ for the corresponding spin-boson model which allows
the observation of a quantum phase transition between a localized
phase and a delocalized phase at a critical $\alpha_{\rm
c}(\Delta)\ge 1$ \cite{Leggett, Weiss}. For the value of $t=0.1$
we find $\alpha_{\rm c}^{Q=-1}\approx 1.2$, naturally independent of
$U$ for the single-electron subspace (dashed line in
Fig.~\ref{fig:pd}a). Note that $\alpha_{\rm c}^{Q=-1}$ deviates
from the exact value  $\alpha_{\rm c}^{Q=-1}(t\to0)$ due to the
finite $t$ {\em and} the value of $\Lambda$ as discussed in
detail in Ref.~\onlinecite{BLTV}.

\begin{figure}[!t]
\epsfxsize=2.8in
\centerline{\epsffile{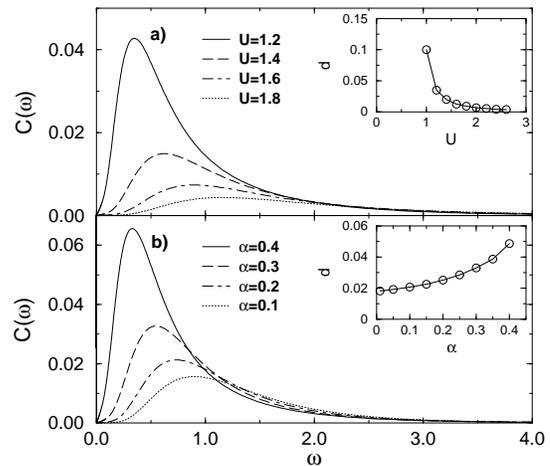}}
\vspace*{-0.4cm}
\caption{Density-density
correlation function $C(\omega)$ as a function of the frequency $\omega$ for
(a) different values of $U$ and fixed $\alpha=0.5$ 
and
(b) different values of $\alpha$ and fixed $U=1.0$ ($t=0.1$ in both
figures). In the insets of (a) and (b),
the double occupancy is displayed as a function of $U$
and $\alpha$, respectively.}
\label{fig:C-and-d}
\end{figure}

In the two-electron subspace we also observe a transition between
a localized and a delocalized phase
(solid line in Fig.~\ref{fig:pd}a). However, the dynamics
of the two electrons is now correlated which leads to a
$U$-dependent $\alpha_{\rm c}^{Q=0} (U)$. Note that even for $U=0$ we
find $\alpha_{\rm c}^{Q=0}(U\!=\!0) \ne \alpha_{\rm c}^{Q=-1}$. This is because
the coupling to the common bosonic bath generates an effective
interaction between the two electrons
[the energy of the total system is lower when both electrons
occupy the same lattice site (D or A) as compared to the case
of equal occupancy on D and A which does not result in a displacement
of the oscillators]. 
In other words,
many-particle effects are present in the system even if the
Coulomb interaction is assumed to be very small which is in close
analogy to the physics of the Holstein model \cite{Holstein}. In our
model, the situation is more complicated; due to the coupling to a
continuous bath, the effective interaction is always frequency
dependent (there is no limit in which it could be reduced to a
static one). Nevertheless, the net sign of the effective
interaction is clearly negative, in agreement with the
behavior of the double occupancy shown in Fig. \ref{fig:C-and-d}
which increases with increasing coupling to the bosonic bath.

This also explains, to some degree, the shape and the asymmetry of the
phase boundary $\alpha_{\rm c}^{Q=0} (U)$: a repulsive $U>0$
partly cancels the attractive interaction from the coupling to the
bosonic bath so that increasing $U$ favors delocalization of
the electrons (for a similar feature in the Hubbard-Holstein model,
see Ref.~\onlinecite{Alex}).

An interesting consequence of this increase
 is that the lines $\alpha_{\rm c}^{Q=-1} (U)$ and
$\alpha_{\rm c}^{Q=0} (U)$ cross at some value of $U^\ast \approx 2.4$;
for $U>U^\ast$ there exists a region in the phase diagram
with  $\alpha_{\rm c}^{Q=-1} (U) < \alpha < \alpha_{\rm c}^{Q=0} (U)$
where the system is localized in the single-electron subspace
but delocalized in the two-electron subspace. This can lead
to interesting effects of the electron transfer dynamics: the
addition of a second electron to the donor site drives the system
from the localized to the delocalized phase so that either one of
the electrons is enabled to hop to the acceptor site
(such a process could be termed Coulomb-assisted transfer)
or both electrons hop in a single event (bipolaronic transfer).

The dependence of the critical $\alpha_{\rm c}$ on $t$ in the
two-electron subspace is shown in Fig.~\ref{fig:pd}b for
various values of $U$. As expected, increasing $t$ tends to delocalize
the electrons, similar to the behavior in the single-electron
subspace (see, for example, Fig.~8 in Ref.~\onlinecite{BLTV}).

To study the dynamics of the electron transfer process in the
two-electron subspace we calculate the density-density
correlation function
$C(\omega)=\frac{1}{2\pi} \int_{-\infty}^{+\infty} e^{i \omega t} C(t)\,
{\rm d}t$ with
\begin{equation}
C(t)=\frac{1}{2} \left \langle  [n_D(t)-n_A(t),n_D(0)-n_A(0)]_+
\right \rangle \ .
\end{equation}
In the single-electron subspace, the quantity corresponds to the
well-studied spin-spin correlation function \cite{Leggett,Weiss}.

Figure \ref{fig:C-and-d} shows the
NRG results for $C(\omega)$ for various values of
$\alpha$ and $U$ with the temperature set to $T\!=\!0$. 
As expected from the corresponding behavior
in the spin-boson model, the slope of $C(\omega)$ for small
$\omega$ diverges upon approaching the transition from the
delocalized side. This can be seen in Fig.~\ref{fig:C-and-d}b where we show
$C(\omega)$ for fixed $U=1.0$ and increasing values of $\alpha <
\alpha_{\rm c}^{Q=0}\approx 0.55$. Such an increasing slope in
 $C(\omega)$ corresponds to a decreasing electron transfer
rate $k$, with $k\to 0$ for $\alpha \to \alpha_{\rm c}^{Q=0}$.

In the insets of Fig.~\ref{fig:C-and-d}, we show  the double occupancy
$d=d_{\rm A}=\langle n_{\rm A\uparrow} n_{\rm A\downarrow}\rangle
= \int_0^\infty C(\omega) {\rm d}
\omega$ (which is equal to $d_{\rm D}$ for zero bias).
The double occupancy increases with increasing $\alpha$; this
is similar to the Holstein model, where the coupling to the phonons
favors the formation of bipolarons due to the attractive
effective interaction. Increasing the value of $U$, on the
other hand, reduces the double occupancy as shown in the inset
of Fig.~\ref{fig:C-and-d}a. As shown in the main panel of
Fig.~\ref{fig:C-and-d}a, increasing
$U$ also leads to an increase of the electron transfer
rate [a decreasing slope in $C(\omega)$].

\begin{figure}[!t]
 \epsfxsize=2.7in
\centerline{\epsffile{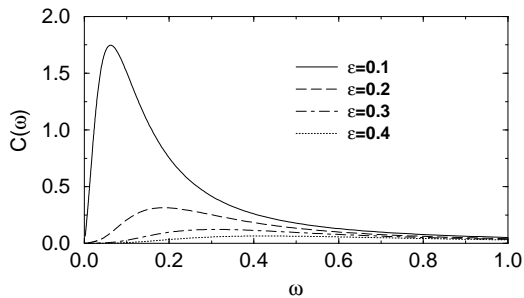}}         
\vspace*{-0.5cm}
\caption{Density-density
correlation function $C(\omega)$ for 
$U=1.0$, $\alpha=0.5$, $t=0.1$, and
different values of the bias $\epsilon$.}
\label{fig:C-epsne0}
\vspace*{-0.5cm}
\end{figure}

As in the spin-boson model, no quantum phase transition 
is observed for any finite $\epsilon$ 
where the system is always in the delocalized phase
\cite{Leggett}. The dependence of the density-density correlation
function on the bias $\epsilon$ is shown
in Fig.~\ref{fig:C-epsne0}. At small frequencies,
we again find $C(\omega)\propto \omega$
with the slope decreasing with increasing $\epsilon$. Here one cannot
deduce directly information on the transfer rate because of its
non-monotonic behavior as a function of $\epsilon$ \cite{Marcus}.

{\em Conclusion} -
To summarize, we have shown that many-particle effects
can lead to significant changes of the ET characteristics in
donor-acceptor systems when more than one electron is present in the
system. In the two-site electron-boson model proposed here,
these many-particle effects are due to both the explicit
inclusion of a local Coulomb interaction $U$ and an effective
interaction induced by the coupling to a common bosonic bath.
We found that, depending on the model parameters, the addition
of a second electron can lead to a transition from localization
to delocalization, or vice versa, see Fig.~\ref{fig:pd}.
The results for the density-density correlation
function and the double occupancy show that increasing the
coupling to the bosonic bath favors the formation of bipolarons, while
increasing the Coulomb repulsion $U$ has the opposite effect.

The bosonic NRG is a very flexible tool and we are planning
to investigate a number of extensions
of the present model in the future. A generalization of the level structure
allows, for example, the modeling of excitation transfer (see
Ref.~\onlinecite{Schulten1}
and the models proposed in Ref.~\onlinecite{Gilmore}). 
Furthermore, the extension
to three or more sites allows the investigation of multistep
transfer processes and the influence of dissipation in the
transport through (short) one-dimensional chains (see, for example,
Ref.~\onlinecite{chain}). In general, the
method presented here
is a starting point for a more realistic description
of electron and excitation transfer in those
molecular systems where many-particle effects have to be considered.

We acknowledge helpful discussions with 
S. Kehrein,
H.-J. Lee, and
D. Vollhardt.
This research was supported by the DFG through SFB 484 (ST,RB),
the Center for Functional Nanostructures (NT), and by
the Alexander von Humboldt foundation (NT).


\end{document}